\documentclass[10pt,conference]{IEEEtran}\IEEEoverridecommandlockouts
\usepackage{epsfig,graphicx,subfigure,psfrag,amsmath,cases,bm}
\usepackage{latexsym,amssymb,amsmath,epsfig,subfigure,algorithm,mathtools}
\usepackage{algorithmic}
\usepackage{color}
\usepackage{url}
\usepackage{scrtime}

\author{\IEEEauthorblockN {Yan Sun\IEEEauthorrefmark {1}, Derrick Wing Kwan Ng\IEEEauthorrefmark {2}, and Robert Schober\IEEEauthorrefmark {1}}

\IEEEauthorrefmark {1}Institute for Digital Communications, Friedrich-Alexander-University
Erlangen-N\"urnberg (FAU), Germany\\
\IEEEauthorrefmark {2}School of Electrical Engineering and Telecommunications, The University
of New South Wales, Australia \vspace*{-4mm}
}

\title{Resource Allocation for MC-NOMA Systems with Cognitive Relaying\vspace*{-2mm}}

\newtheorem{T-Prob}{Transformed Problem}

\DeclareMathOperator{\Tr}{Tr}

\DeclareMathOperator{\maxo}{maximize}
\DeclareMathOperator{\mino}{minimize}

\newtheorem{Remark}{Remark}

\newcommand{\abs}[1]{\lvert#1\rvert}

\begin{document}
%\IEEEspecialpapernotice{(Invited Paper)}
\maketitle

\begin{abstract}
In this paper, we investigate the resource allocation algorithm design for cooperative cognitive relaying multicarrier non-orthogonal multiple access (MC-NOMA) systems.
In particular, the secondary base station serves multiple secondary users and simultaneously acts as a relay assisting the information transmission in the primary network.
The resource allocation aims to maximize the weighted system throughput by jointly optimizing the power and subcarrier allocation for both the primary and the secondary networks while satisfying the quality-of-service requirements of the primary users.
The algorithm design is formulated as a mixed combinatorial non-convex optimization problem.
We apply monotonic optimization theory to solve the problem leading to an optimal resource allocation policy.
Besides, we develop a low-complexity scheme to find a suboptimal solution.
Our simulation results reveal that the performance of the proposed suboptimal algorithm closely  approaches that of the optimal one.
Besides, the combination of MC-NOMA and cognitive relaying improves the system throughput  considerably compared to conventional multicarrier cognitive relaying systems.
\end{abstract}

\renewcommand{\baselinestretch}{0.98}
\large\normalsize

\vspace*{-3mm}
\section{Introduction}
Spectrum has become a scarce resource due to the continuously growing demand for high-data rate communications. This has created a bottleneck for providing ubiquitous communication services.
To handle this issue, cognitive radio (CR) has been proposed as a promising technique to improve spectrum utilization by enabling an unlicensed network (e.g. the secondary network) to dynamically access the licensed spectrum of the primary network \cite{Chen2014SurveyCR}.
However, the deployment of CR degrades the performance of the primary network due to the co-channel interference originating from the secondary network \cite{sun2016robustCR}.
Recently, cooperative CR has attracted significant research interest since it can reduce the performance degradation to the primary network caused by CR deployment.
In particular, in cooperative CR networks, the secondary base station (BS) acts as a relay to assist the signal transmission of the primary BS, while simultaneously utilizing the licensed spectrum to serve the secondary users (SUs).
In \cite{Han09CooperativeRelay}, the authors studied the outage probability of cooperative CR relaying systems.
The authors of \cite{Manna11CoopRelayMIMO} focused on  precoder design for  cooperative CR systems where a multiple-antenna secondary BS was deployed for suppressing the interference to primary users (PUs).
However, since the primary network and the secondary network coexist in the same frequency band, there is a nontrivial tradeoff between the performance of the two networks due to the mutual interference.

Recently, non-orthogonal multiple access (NOMA) has been proposed to improve spectral efficiency by harnessing co-channel interference via superposition coding at the transmitter and successive interference cancellation (SIC) at the receiver \cite{book:Key5GWong}--\nocite{Wei17optimal}\cite{sun2016optimalJournal}.
Specifically, NOMA multiplexes the message of multiple users on the same time-frequency resource and exploits the power domain for multiple access.
Motivated by this, the application of NOMA in CR systems for improving the spectral efficiency was  investigated in \cite{ImpactPairNOMA}, where the secondary users were equipped with successive interference cancellers to cancel the interference generated by the primary network.
Subsequently, the concept of NOMA was extended to cognitive relaying systems \cite{Lv2017NOMAcooper,Lv2017Design}.
For instance, in \cite{Lv2017NOMAcooper}, the outage probability of a NOMA-based cognitive relaying system was investigated with the goal to satisfy the quality-of-service (QoS) requirements of the PU.
The authors of \cite{Lv2017Design}  studied the performance of a cognitive NOMA system where a group of multicast SUs relayed  information to a single PU.
However, only fixed power allocation was adopted in \cite{Lv2017NOMAcooper,Lv2017Design} to facilitate the performance analysis.
However, power allocation is critical for improving the performance of CR systems and the optimal resource allocation design for NOMA with cognitive relaying is still unknown.
Moreover, \cite{Lv2017NOMAcooper,Lv2017Design}  studied only single-carrier systems for serving a single PU, thus the optimal resource allocation design for multiuser cognitive relaying multicarrier NOMA (MC-NOMA) systems is still an open problem.

In this paper, we address the above issues. To this end, the resource allocation algorithm design for cognitive relaying MC-NOMA systems is formulated as a non-convex optimization problem for the maximization of the weighted system throughput. We solve the considered problem optimally via monotonic optimization theory \cite{tuy2000monotonic,zhang2013monotonic} and obtain the optimal power and subcarrier allocation policy. In addition, we also develop a low-complexity suboptimal scheme which achieves a close-to-optimal performance.

\vspace*{-1mm}
\section{System Model}
In this section, we present the adopted notation and the considered cognitive relaying MC-NOMA system model.

\vspace*{-1mm}
\subsection{Notation}%
We use boldface capital and lower case letters to denote matrices and vectors, respectively. $\mathbf{a}^T$ denotes the transpose of vector $\mathbf{a}$; $\Tr(\mathbf{A})$ denotes the trace of matrix $\mathbf{A}$; $\mathbb{C}$ denotes the set of complex numbers; $\mathbb{R}^+$ denotes the set of non-negative real numbers;  $\abs{\cdot}$ denotes the absolute value of a complex scalar; ${\cal E}\{\cdot\}$ denotes statistical expectation. The circularly symmetric complex Gaussian distribution with mean $w$ and variance $\sigma^2$ is denoted by ${\cal CN}(w,\sigma^2)$; and $\sim$ stands for ``distributed as". $\nabla_{\mathbf{x}} f(\mathbf{x})$ denotes the gradient vector of function $f(\mathbf{x})$ whose components are the partial derivatives of $f(\mathbf{x})$.

\vspace*{-1mm}
\subsection{Cognitive Relaying MC-NOMA System Model}%
\begin{figure}
\centering\vspace*{-0mm}
\includegraphics[width=3in]{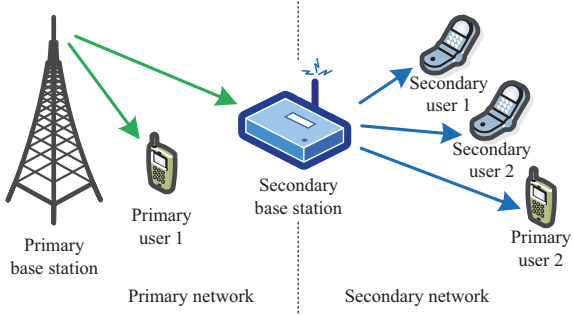}\vspace*{-3mm}
\caption{A cognitive relaying system with a primary base station, $K=2$ primary users, one secondary BS, and $J=2$ secondary users.}
\label{fig:system_model}\vspace*{-5mm}
\end{figure}
\begin{figure}
\centering\vspace*{-0mm}
\includegraphics[width=3in]{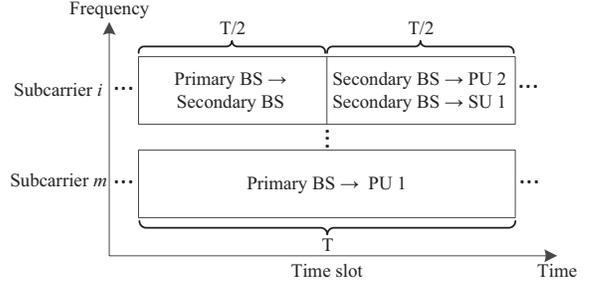}\vspace*{-3mm}
\caption{Illustration of the subcarrier allocation in one time slot.}
\label{fig:scheduling}\vspace*{-5mm}
\end{figure}

The considered cognitive relaying MC-NOMA system comprises one primary BS, $K$ PUs, one secondary BS, and $J$ SUs. All transceivers are single-antenna half-duplex (HD) devices, cf. Figure \ref{fig:system_model}. The entire frequency band of $W$ Hertz is partitioned into $N_{\mathrm{F}}$ orthogonal subcarriers. The duration of a time slot is $T$.
The primary BS serves the PUs in the downlink (DL) in each scheduling slot. For the case that a PU is far away from the primary BS or has a poor channel to the primary BS, the secondary BS can act as a decode-and-forward (DF) relay \cite{zhangperformance} to assist the primary BS by forwarding the information received from the primary BS to the intended PU. For example, assuming that the secondary BS assists the transmission of PU $k$, the primary BS transmits the information of PU $k$ to the secondary BS in the first half of a time slot on subcarrier $i$. Then, during the second half of the time slot, the secondary BS decodes the received signal on subcarrier $i$ and forwards the recovered information to PU $k$ on the same subcarrier, cf. Figure \ref{fig:scheduling}.
Meanwhile, the secondary BS also serves the SUs via the subcarriers that are used for forwarding information to the PUs in the second half of the time slot.
We assume that the secondary BS adopts NOMA to provide wireless service to the SUs by multiplexing the message of one secondary user (SU) and one PU on the same subcarrier.
In addition, we assume that the SUs are equipped with successive interference cancellers for multiuser detection but the PUs are only equipped with linear receivers for single-user detection. Thus, if a PU is assisted by the secondary BS, it will be subject to the co-channel interference originating from the signal of the SU that is served on the same subcarrier.
\begin{Remark}
We assume that the secondary BS can multiplex at most one SU and one PU on each subcarrier\footnote{Multiplexing more than two users on a subcarrier leads to more severe co-channel interference and requires several stages of SIC at the users which increases the complexity and introduces additional delays.}. For the primary network, each subcarrier can only be allocated to at most one PU or to the secondary BS for information relaying\footnote{The primary BS is assumed to be a legacy infrastructure which adopts conventional orthogonal multiple access.}.
\end{Remark}
\vspace*{-1mm}
\subsection{Channel Model}%
For the primary network, the primary BS employs orthogonal multiple access which indicates that each subcarrier can either be allocated for transmitting the signal to one PU or to the secondary BS. Specifically, for a given time slot, if PU $k\in\{1,\ldots,K\}$ is scheduled on subcarrier $i\in\{1,\ldots,N_{\mathrm{F}}\}$, the received signal at PU $k$ on subcarrier $i$ is given by \vspace*{-2mm}
\begin{eqnarray}
x_k^i=\sqrt{q_k^i}f_k^i d_{\mathrm{PU}_k}^i + z_{\mathrm{PU}_k}^i,
\end{eqnarray}
where $d_{\mathrm{PU}_k}^i\in\mathbb{C}$ denotes the information symbol intended for PU $k$ on subcarrier $i$ and we assume ${\cal E}\{\abs{d_{\mathrm{PU}_k}^i}^2\}=1$ without loss of generality.
$q_k^i\in\mathbb{R}^+$ is the transmit power for the signal transmitted directly to PU $k$.
$f_k^i\in\mathbb{C}$ denotes the channel coefficient of the link between the primary BS and PU $k$ on subcarrier $i$ and captures the joint effect of pathloss, small scale fading, and shadowing. $z_{\mathrm{PU}_k}^i\sim{\cal CN}(0,\sigma_{\mathrm{PU}_k}^2)$ denotes the complex additive white Gaussian noise (AWGN) on subcarrier $i$ at PU $k$.
Besides, if subcarrier $i$ is allocated for transmitting the information of PU $k$ via the secondary BS, the received signal at the secondary BS on subcarrier $i$ in the first half of the time slot is given by \vspace*{-2mm}
\begin{eqnarray}
x_{\mathrm{ST}}^i=\sqrt{q_{\mathrm{ST}}^i}f_{\mathrm{ST}}^i d_{\mathrm{PU}_k}^i + z_{\mathrm{ST}}^i,
\end{eqnarray}
where $q_{\mathrm{ST}}^i\in\mathbb{R}^+$ is the transmit power for the signal transmitted to the assisting secondary BS. $f_{\mathrm{ST}}^i\in\mathbb{C}$ denotes the channel coefficient between the primary BS and the secondary BS on subcarrier $i$. $z_{\mathrm{ST}}^i\sim{\cal CN}(0,\sigma_{\mathrm{ST}}^2)$ denotes the complex AWGN on subcarrier $i$ at the secondary BS.

On the other hand, in the second half of the time slot, the secondary BS can multiplex at most one PU and one SU on each subcarrier in the secondary network. Specifically, assuming that PU $k$ and SU $j\in\{1,\ldots,J\}$ are multiplexed on subcarrier $i$, the received signals at PU $k$ and SU $j$ are given by \vspace*{-2mm}
\begin{eqnarray}
\hspace*{-10mm}&& y_{\mathrm{PU}_k}^i=\sqrt{p_{\mathrm{PU}_k}^i} h_k^i d_{\mathrm{PU}_k}^i + \sqrt{p_{\mathrm{SU}_j}^i} h_k^i d_{\mathrm{SU}_j}^i + z_{\mathrm{PU}_k}^i  \,\,\,\, \text{and} \\
\hspace*{-10mm}&& y_{\mathrm{SU}_j}^i=\sqrt{p_{\mathrm{SU}_j}^i}g_j^i d_{\mathrm{SU}_j}^i + \sqrt{p_{\mathrm{PU}_k}^i}g_j^i d_{\mathrm{PU}_k}^i + z_{\mathrm{SU}_j}^i,
\end{eqnarray}
respectively, where $d_{\mathrm{SU}_j}^i\in\mathbb{C}$ denotes the information symbol intended for SU $j$ on subcarrier $i$, and we assume ${\cal E}\{\abs{d_{\mathrm{SU}_j}^i}^2\}=1$ without loss of generality.  $p_{\mathrm{PU}_k}^i\in\mathbb{R}^+$ and $p_{\mathrm{SU}_j}^i\in\mathbb{R}^+$ denote the transmit powers for the signals intended for PU $k$ and SU $j$ on subcarrier $i$ at the secondary BS, respectively.
$h_k^i\in\mathbb{C}$ and $g_j^i\in\mathbb{C}$ denote the coefficients for the secondary BS-to-PU $k$ link and the secondary BS-to-SU $j$ link on subcarrier $i$, respectively. We note that the joint effect of pathloss, small scale fading, and shadowing is captured by variables $h_k^i$ and $g_j^i$. $z_{\mathrm{SU}_j}^i\sim{\cal CN}(0,\sigma_{\mathrm{SU}_j}^2)$ denotes the complex AWGN on subcarrier $i$ at SU $j$.
Besides, for the study of optimal resource allocation algorithm design, we assume that the global channel state information (CSI) of all links is perfectly known at the primary and secondary BSs.

\vspace*{-1mm}
\section{Problem Formulation}
In this section, we first define the adopted performance measure for the considered cognitive relaying MC-NOMA system. Then, we formulate the power and subcarrier allocation design as an optimization problem.

\vspace*{-1mm}
\subsection{Achievable Rate and Weighted System Throughput}
In the primary network, if subcarrier $i$ is allocated to PU $k$ for direct transmission, the achievable data rate of PU $k$ on subcarrier $i$ is given by \vspace*{-2mm}
\begin{eqnarray}
C_{\mathrm{PU}_{k}}^i = \log_2 \Big( 1 + q_{k}^i F_k^i\Big),
\end{eqnarray}
where $F_k^i = \abs{f_k^i}^2/\sigma_{\mathrm{PU}_k}^2$; otherwise if the secondary BS assists the information transmission to a PU on subcarrier $i$, the achievable data rate of the secondary BS on subcarrier $i$ is given by \vspace*{-2mm}
\begin{eqnarray}\label{rate_PT2ST}
C_{\mathrm{ST}}^i = \frac{1}{2}  \log_2 \Big( 1 + q_{\mathrm{ST}}^i F_{\mathrm{ST}}^i\Big),
\end{eqnarray}
where $F_{\mathrm{ST}}^i = \abs{f_{\mathrm{ST}}^i}^2/\sigma_{\mathrm{ST}}^2$ and the pre-log factor $1/2$ appears in \eqref{rate_PT2ST} since the primary BS communicates with the secondary BS only in the first half of the time slot.

On the other hand, the secondary network adopts NOMA for improving the system performance.
Since only SUs are equipped with successive interference cancellers, we adopt a fixed SIC decoding order to enable NOMA.
Specifically, assuming that PU $k$ and SU $j$ are multiplexed simultaneously on subcarrier $i$ in the second half of the time slot, SU $j$ performs SIC to decode and remove the signal of PU $k$ before decoding its own signal. Then, the achievable data rates of PU $k$ and SU $j$ are given by \vspace*{-2mm}
\begin{eqnarray}
\label{rate_ST2PU}
\hspace*{-10mm} R_{\mathrm{PU}_{k,j}}^i \hspace*{-2mm} &=& \hspace*{-2mm} \frac{1}{2} \log_2 \Big( 1 + \frac{p_{\mathrm{PU}_k}^i H_k^i}{p_{\mathrm{SU}_j}^i H_k^i + 1}\Big) \,\,\,\, \text{and} \\[-1mm]
%%%%
\label{rate_ST2SU}
\hspace*{-10mm} R_{\mathrm{SU}_{k,j}}^i \hspace*{-2mm} &=& \hspace*{-2mm} \frac{1}{2}  \log_2 (1 + p_{\mathrm{SU}_j}^i  G_j^i),
\end{eqnarray}
respectively, where $H_k^i = \abs{h_k^i}^2/\sigma_{\mathrm{PU}_k}^2$ and $G_j^i = \abs{g_j^i}^2/\sigma_{\mathrm{SU}_j}^2$. The achievable rates in \eqref{rate_ST2PU} and \eqref{rate_ST2SU} are multiplied with a factor of $1/2$ since the secondary BS utilizes only the second half of the time slot for transmitting signals to the SUs and PUs. Besides, the following inequality specifies the maximum achievable data rate for relaying the information of PU $k$ on subcarrier $i$ in the secondary network: \vspace*{-2mm}
\begin{eqnarray} \label{relay-ineq}
s_{k,j}^i R_{\mathrm{PU}_{k,j}}^i \le c_{\mathrm{ST}}^i C_{\mathrm{ST}}^i, \forall i,k,j,
\end{eqnarray}
where $s_{k,j}^i \in \{1,0\}$ and $c_{\mathrm{ST}}^i \in \{1,0\}$ are the binary subcarrier allocation indicators. In particular, $s_{k,j}^i=1$ indicates that PU $k$ and SU $j$ are multiplexed on subcarrier $i$ in the secondary network, $c_{\mathrm{ST}}^i=1$ indicates that subcarrier $i$ is allocated for transmitting information from the primary BS to the secondary BS in the primary network, and $s_{k,j}^i=0$ and $c_{\mathrm{ST}}^i=0$ if subcarrier $i$ is not used due to unsatisfactory channel conditions.

Moreover, the use of NOMA requires successful SIC at the SU for interference mitigation. In practice, SU $j$ can successfully decode and remove the co-channel interference causing by PU $k$ on subcarrier $i$ by SIC only when the following inequality holds:\vspace*{-2mm}
\begin{eqnarray} \label{SIC-inequality}
 \hspace*{-4mm} \log_2 \Big( 1 + \frac{p_{\mathrm{PU}_k}^i H_k^i}{p_{\mathrm{SU}_j}^i H_k^i + 1}\Big) \hspace*{-2mm} &\le& \hspace*{-2mm} \log_2 \Big( 1 + \frac{p_{\mathrm{PU}_k}^i G_j^i}{p_{\mathrm{SU}_j}^i G_j^i + 1}\Big) \\
\label{simple-SIC-inequality}
 \hspace*{-4mm} \Longleftrightarrow \hspace*{15mm} H_k^i -  G_j^i\hspace*{-2mm} &\le& \hspace*{-2mm} 0.
\end{eqnarray}
We note that the inequality in \eqref{SIC-inequality} is due to the fixed decoding order in the considered system.
Besides, the inequality in \eqref{simple-SIC-inequality} indicates that for a given subcarrier $i$, an SU can only be paired with a PU that has a worse DL transmission channel with respect to the secondary BS than the SU. Therefore, we define $\mathcal{K}(i,j)$ as the set of PUs whose channel condition satisfies \eqref{simple-SIC-inequality}, i.e., $H_k^i  -  G_j^i  \le  0, \forall k \in \mathcal{K}(i,j)$.

Therefore, the weighted system throughput on subcarrier $i$ is given by\vspace*{-2mm}
\begin{eqnarray} \label{throughput-i}
\hspace*{0mm}U^i \hspace*{-1mm}
=\hspace*{-1mm}  \sum_{j=1}^{J}   \sum_{k\in \mathcal{K}(i,j)} \hspace*{-3mm} s_{k,j}^i \Big[w R_{\mathrm{PU}_{k,j}}^i \hspace*{-1mm} + \hspace*{-1mm} \mu  R_{\mathrm{SU}_{k,j}}^i \Big] \hspace*{-1mm} + \hspace*{-1mm} \sum_{k=1}^{K} c_k^i w C_{\mathrm{PU}_{k}}^i.
\end{eqnarray}
Here, $c_{k}^i \in \{1,0\}$ and $c_{k}^i=1$ indicates that PU $k$ is scheduled on subcarrier $i$ in the primary network, otherwise $c_{k}^i=0$. The positive constants $0 \le w \le 1$ and $0 \le \mu \le 1$ reflect the priority of the PUs and SUs in resource allocation, respectively, and are specified in the media access control (MAC) layer to achieve certain fairness objectives in resource allocation.

\vspace*{-1mm}
\subsection{Optimization Problem Formulation}%
The design objective is to maximize the weighted sum throughput of the two systems, while guaranteeing minimum data rates for the PUs. The optimal joint power and subcarrier allocation policy is obtained by solving the following optimization problem:\vspace*{-2mm}
\begin{eqnarray} \label{pro}
&&\hspace*{-7mm}\underset{ \substack{{q_k^i, q_{\mathrm{ST}}^i, p_{\mathrm{PU}_k}^i, p_{\mathrm{SU}_j}^i \ge 0,}\\{c_{\mathrm{ST}}^i, c_{k}^i, s_{k,j}^i}}}{\maxo}\,\, \,\,  \sum_{i=1}^{N_{\mathrm{F}}} U^i  \\  [-1mm]
\notag\mbox{s.t.}
%%%%%
&&\hspace*{-5mm}\mbox{C1: } s_{k,j}^i R_{\mathrm{PU}_{k,j}}^i - c_{\mathrm{ST}}^i C_{\mathrm{ST}}^i \le 0, \forall i,k,j,\notag  \\ [-1mm]
%%%%%
&&\hspace*{-5mm}\mbox{C2: } \overset{N_{\mathrm{F}}}{\underset{i=1}{\sum}}\Big(   c_k^i C_{\mathrm{PU}_{k}}^i + \overset{J}{\underset{j=1}{\sum}} s_{k,j}^i R_{\mathrm{PU}_{k,j}}^i \Big) \ge R_{{\mathrm{PU}}_k}^{\mathrm{req}},\,\, \forall k, \notag\\ [-1mm]
%%%%%%
&&\hspace*{-5mm}\mbox{C3: } \overset{N_{\mathrm{F}}}{\underset{i=1}{\sum}}  \overset{K}{\underset{k=1}{\sum}}  \overset{J}{\underset{j=1}{\sum}} s_{k,j}^i (p_{\mathrm{PU}_k}^i + p_{\mathrm{SU}_j}^i ) \le P_{\mathrm{max}}^{\mathrm{ST}}, \notag \\[-1mm]
%%%%%%
&&\hspace*{-5mm}\mbox{C4: } \overset{N_{\mathrm{F}}}{\underset{i=1}{\sum}} \Big(c_{\mathrm{ST}}^i q_{\mathrm{ST}}^i + \overset{K}{\underset{k=1}{\sum}} c_k^i q_{k}^i \Big)\le P_{\mathrm{max}}^{\mathrm{PT}}, \notag \\[-1mm]
%%%%
&&\hspace*{-5mm}\mbox{C5: } s_{k,j}^i \in \{0,1\}, \forall i,k,j, \quad \, \mbox{C6: } c_{\mathrm{ST}}^i, c_{k}^i \in \{0,1\}, \forall i,k, \notag\\[-1mm]
%%%%%
&&\hspace*{-5mm}\mbox{C7: } c_{k}^i + \overset{J}{\underset{j=1}{\sum}} s_{k,j}^i \le 1, \forall i,k, \,\, \mbox{C8: }  c_{\mathrm{ST}}^i + \overset{K}{\underset{k=1}{\sum}} c_{k}^i \le 1, \forall i. \notag
\end{eqnarray}
In problem \eqref{pro}, constraint C1 guarantees successful information decoding on subcarrier $i$ at PU $k$ if the secondary BS is selected for assisting information transmission. Constraint C2 imposes a minimum data rate requirement $R_{{\mathrm{PU}}_k}^{\mathrm{req}}$ for PU $k$.
Constraints C3 and C4 are the power budget constraints for the secondary and primary BSs with the maximum transmit powers $P_{\mathrm{max}}^{\mathrm{ST}}$ and $P_{\mathrm{max}}^{\mathrm{PT}}$, respectively.
Constraints C5, C6, and C7 are imposed to guarantee that each subcarrier can only be allocated to at most one SU and one PU in the secondary network. Constraints C6 and C8 guarantee that each subcarrier can only be allocated to at most one PU in the primary network or forward the  information of one PU to the secondary BS.

The problem in \eqref{pro} is a mixed-integer non-convex optimization problem which is known to be NP-hard. There exists no systematic and computationally efficient approach for obtaining a globally optimal solution for such problems. Nevertheless, by exploiting the special structure of the considered problem, we will develop efficient resource allocation algorithms for finding the optimal and suboptimal resource allocation policies for \eqref{pro} in the next section.

\vspace*{-1mm}
\section{Solutions of the Optimization Problem}%
In this section, we first solve problem \eqref{pro} optimally by applying monotonic optimization theory \cite{tuy2000monotonic,zhang2013monotonic}. Then, we propose a suboptimal algorithm with low computational complexity for obtaining a close-to-optimal solution of \eqref{pro}.

\vspace*{-0mm}
\subsection{Optimal Resource Allocation Scheme} \label{Optimal-section}
First, we can rewrite the weighted system throughput on subcarrier $i$ in the following equivalent form:\vspace*{-2mm}
\begin{eqnarray} \label{throughput-i-interference}
\hspace*{-6mm}U^i\hspace*{-3mm}
%%%%%%%%%%%%%%%%%%%
 &{=}& \hspace*{-3mm} \sum_{j=1}^{J}   \sum_{k\in \mathcal{K}(i,j)} \hspace*{-1mm}  \frac{1}{2} \Big[w  \log_2 \Big( 1 \hspace*{-0.7mm} + \hspace*{-0.7mm} \frac{s_{k,j}^i p_{\mathrm{PU}_k}^i H_k^i}{s_{k,j}^i p_{\mathrm{SU}_j}^i H_k^i
\hspace*{-0.7mm} + \hspace*{-0.7mm} I_{\mathrm{PU}_{k,j}}^i  \hspace*{-0.7mm} + \hspace*{-0.7mm} 1}\Big) \notag \\[-1mm]
%%%%%
\hspace*{-4mm} &+& \hspace*{-3mm} \hspace*{-0.7mm} \mu  \hspace*{-0.4mm}  \log_2 \hspace*{-1mm} \Big( \hspace*{-0.7mm} 1 \hspace*{-0.7mm} + \hspace*{-0.7mm} \frac{s_{k,j}^i p_{\mathrm{SU}_j}^i \hspace*{-0.7mm} G_j^i}{I_{\mathrm{SU}_{k,j}}^i \hspace*{-0.7mm} + \hspace*{-0.7mm} 1}\Big) \Big] \hspace*{-0.7mm} + \hspace*{-0.7mm} \sum_{k=1}^{K} \hspace*{-0.7mm} w \log_2 \hspace*{-0.5mm} \Big( 1 \hspace*{-0.7mm} + \hspace*{-0.7mm} \frac{c_k^i q_{k}^i F_k^i}{S_{\mathrm{PU}_{k}}^i \hspace*{-0.7mm} + \hspace*{-0.7mm} 1}\Big),
\end{eqnarray}
where\vspace*{-3mm}
\begin{eqnarray}
I_{\mathrm{PU}_{k,j}}^i \hspace*{-2mm} &=& \hspace*{-2mm} \Big( c_k^i q_k^i + \overset{J}{\underset{n \neq j}{\sum}}   \overset{K}{\underset{m \neq k}{\sum}}s_{m,n}^i (p_{\mathrm{PU}_m}^i + p_{\mathrm{SU}_n}^i) \Big)H_k^i, \,\,\,\,\,\,  \\[-2mm]
%%%%
I_{\mathrm{SU}_{k,j}}^i \hspace*{-2mm} &=& \hspace*{-2mm} \Big( c_k^i q_k^i + \overset{J}{\underset{n \neq j}{\sum}}   \overset{K}{\underset{m \neq k}{\sum}}s_{m,n}^i (p_{\mathrm{PU}_m}^i + p_{\mathrm{SU}_n}^i) \Big)G_j^i, \,\,\,\,\,\,  \\[-2mm]
%%%%
S_{\mathrm{PU}_{k}}^i \hspace*{-2mm} &=& \hspace*{-2mm} \Big(c_{\mathrm{ST}}^i q_{\mathrm{ST}}^i + \overset{K}{\underset{m \neq k}{\sum}} c_m^i q_m^i\Big) F_k^i.
\end{eqnarray}
We note that \eqref{throughput-i-interference} is equivalent to \eqref{throughput-i} due to constraints C5--C8. In particular, $I_{\mathrm{PU}_{k,j}}^i$, $I_{\mathrm{SU}_{k,j}}^i$, and $S_{\mathrm{PU}_{k}}^i$ represent the undesired interference at the receivers. Specifically, if a given subcarrier allocation policy satisfies constraints C5--C8, $I_{\mathrm{PU}_{k,j}}^i$, $I_{\mathrm{SU}_{k,j}}^i$, and $S_{\mathrm{PU}_{k}}^i$ do not have any impact on the system performance, otherwise they will severely degrade the system throughput.
In other words, $I_{\mathrm{PU}_{k,j}}^i$, $I_{\mathrm{SU}_{k,j}}^i$, and $S_{\mathrm{PU}_{k}}^i$ act as penalty terms to penalize the objective function for any violation of constraints C5--C8.

Then, we define $\tilde{p}_{\mathrm{PU}_{k,j}}^i = s_{k,j}^i p_{\mathrm{PU}_k}^i$, $\tilde{p}_{\mathrm{SU}_{k,j}}^i = s_{k,j}^i p_{\mathrm{SU}_j}^i$, $\tilde{q}_{k}^i = c_k^i q_{k}^i$, and $\tilde{q}_{\mathrm{ST}}^i = c_{\mathrm{ST}}^i q_{\mathrm{ST}}^i$. Therefore, the weighted system throughout in \eqref{throughput-i-interference} can be rewritten in equivalent form as:\vspace*{-2mm}
\begin{eqnarray} \label{throughput-i-newdefine}
\hspace*{-6mm} \tilde{U}^i\hspace*{-3mm}
&=& \hspace*{-3mm}   \sum_{j=1}^{J}   \sum_{k\in \mathcal{K}(i,j)} \hspace*{-1mm}  \frac{1}{2} \Big[w  \log_2 \Big( 1 \hspace*{-0.7mm} + \hspace*{-0.7mm} \frac{\tilde{p}_{\mathrm{PU}_{k,j}}^i H_k^i}{ \tilde{p}_{\mathrm{SU}_{k,j}}^i H_k^i
\hspace*{-0.7mm} + \hspace*{-0.7mm} \tilde{I}_{\mathrm{PU}_{k,j}}^i  \hspace*{-0.7mm} + \hspace*{-0.7mm} 1}\Big) \notag \\[-1mm]
%%%%%
\hspace*{-6mm} &+& \hspace*{-3mm} \hspace*{-0.7mm} \mu  \hspace*{-0.4mm}  \log_2 \Big(1 \hspace*{-0.7mm} + \hspace*{-0.7mm} \frac{\tilde{p}_{\mathrm{SU}_{k,j}}^i \hspace*{-0.7mm} G_j^i}{\tilde{I}_{\mathrm{SU}_{k,j}}^i \hspace*{-0.7mm} + \hspace*{-0.7mm} 1}\Big) \Big] \hspace*{-0.7mm} + \hspace*{-0.7mm} \sum_{k=1}^{K} \hspace*{-0.7mm} w \log_2 \Big( 1 \hspace*{-0.7mm} + \hspace*{-0.7mm} \frac{\tilde{q}_{k}^i F_k^i}{\tilde{S}_{\mathrm{PU}_{k}}^i \hspace*{-0.7mm} + \hspace*{-0.7mm} 1}\Big),
%%%%
\end{eqnarray}
where
%%%%%
$\tilde{I}_{\mathrm{PU}_{k,j}}^i \hspace*{-1mm}= \hspace*{-1mm}\Big( \tilde{q}_{k}^i \hspace*{-1mm}+ \hspace*{-1mm}\overset{J}{\underset{n \neq j}{\sum}}   \overset{K}{\underset{m \neq k}{\sum}} \tilde{p}_{\mathrm{PU}_{m,n}}^i \hspace*{-1mm} + \hspace*{-0.2mm} \tilde{p}_{\mathrm{SU}_{m,n}}^i \Big)H_k^i$,
%%%%
$\tilde{I}_{\mathrm{SU}_{k,j}}^i \hspace*{-4mm} = \hspace*{-1mm} \Big( \tilde{q}_{k}^i \hspace*{-1mm} + \hspace*{-2mm} \overset{J}{\underset{n \neq j}{\sum}} \hspace*{-0.5mm}  \overset{K}{\underset{m \neq k}{\sum}} \tilde{p}_{\mathrm{PU}_{m,n}}^i \hspace*{-1.5mm} + \hspace*{-0.5mm} \tilde{p}_{\mathrm{SU}_{m,n}}^i \Big)G_j^i$,
%%%%
and $\tilde{S}_{\mathrm{PU}_{k}}^i \hspace*{-1mm} = \hspace*{-1mm} \Big(\tilde{q}_{\mathrm{ST}}^i \hspace*{-0.5mm} + \hspace*{-2mm} \overset{K}{\underset{m \neq k}{\sum}}  \hspace*{-1mm} \tilde{q}_{m}^i \Big) F_k^i$.
Besides, we note that non-convex constraint C1 is the difference of two logarithmic functions and not monotonic. Hence, constraint C1 is not in the canonical form necessary for applying monotonic optimization. Thus, with the aforementioned definitions, we handle the non-convex constraint C1 by equivalently expressing it as\vspace*{-2mm}
\begin{eqnarray}
&&\hspace*{-13mm}\mbox{C1a: }
\hspace*{-1mm} \log_2 \hspace*{-1mm}\Big( \hspace*{-0.5mm} 1 \hspace*{-1mm} + \hspace*{-1mm} \frac{\tilde{p}_{\mathrm{PU}_{k,j}}^i  \hspace*{-0.7mm} H_k^i}{ \tilde{p}_{\mathrm{SU}_{k,j}}^i \hspace*{-1mm} H_k^i
\hspace*{-0.7mm} + \hspace*{-1mm} \tilde{I}_{\mathrm{PU}_{k,j}}^i  \hspace*{-2mm} + \hspace*{-1.5mm} 1} \hspace*{-0.5mm} \Big) \hspace*{-1mm} + \hspace*{-1mm} \varsigma_{k,j}^i  \hspace*{-1mm} \le \hspace*{-1mm} \log_2 \hspace*{-1mm} \Big (\hspace*{-0.7mm} 1 \hspace*{-1mm} + \hspace*{-1mm} P_{\mathrm{max}}^{\mathrm{ST}}H_k^i \hspace*{-0.5mm} \Big) \hspace*{-0.7mm},  \\
%%%%%
%%%%
&&\hspace*{-13mm}\mbox{C1b: }
\hspace*{-1.3mm} \log_2 \hspace*{-0.7mm} \Big( 1 + \tilde{q}_{\mathrm{ST}}^i F_{\mathrm{ST}}^i\Big) \hspace*{-0mm} + \hspace*{-0mm} \varsigma_{k,j}^i   \hspace*{-0mm} \ge \hspace*{-0mm} \log_2 \hspace*{-0.7mm} \Big ( \hspace*{-0.2mm} 1 \hspace*{-0mm} + \hspace*{-0mm} P_{\mathrm{max}}^{\mathrm{ST}}H_k^i \hspace*{-0.5mm} \Big),
\end{eqnarray}
where $ \varsigma_{k,j}^i \ge 0$ is a slack scalar optimization variable. We note that constraints  $\mbox{C1}\mbox{a}$ and $\mbox{C1b}$ are monotonically increasing functions.

Then, the original problem in \eqref{pro} can be rewritten in the following equivalent form: \vspace*{-2mm}
\begin{eqnarray} \label{pro-trans}
&&\hspace*{-7mm}\underset{ \tilde{q}_k^i, \tilde{q}_{\mathrm{ST}}^i, \tilde{p}_{\mathrm{PU}_{k,j}}^i, \tilde{p}_{\mathrm{SU}_{k,j}}^i}{\maxo}\,\, \,\,  \sum_{i=1}^{N_{\mathrm{F}}} \tilde{U}^i   \\  [-1mm]
\notag\mbox{s.t.}
%%%%%
&&\hspace*{-5mm}\mbox{C1}\mbox{a}, \mbox{C1b}, \quad \mbox{C3: } \overset{N_{\mathrm{F}}}{\underset{i=1}{\sum}}  \overset{K}{\underset{k=1}{\sum}}  \overset{J}{\underset{j=1}{\sum}} \,\, \tilde{p}_{\mathrm{PU}_{k,j}}^i \hspace*{-1mm} + \hspace*{-0.5mm} \tilde{p}_{\mathrm{SU}_{k,j}}^i  \le P_{\mathrm{max}}^{\mathrm{ST}}, \notag \\[-1mm]
%%%%%
&&\hspace*{-5mm}\mbox{C2: } \overset{N_{\mathrm{F}}}{\underset{i=1}{\sum}}\Big( \hspace*{-0.7mm} \log_2 ( 1 \hspace*{-0.7mm} + \hspace*{-0.7mm} \tilde{q}_{k}^i F_k^i ) \hspace*{-0.7mm} + \hspace*{-0.7mm} \overset{J}{\underset{j=1}{\sum}}  \frac{1}{2} \log_2 \Big( 1 \hspace*{-0.7mm} + \hspace*{-0.7mm} \frac{\tilde{p}_{\mathrm{PU}_{k,j}}^i H_k^i}{\tilde{p}_{\mathrm{SU}_{k,j}}^i H_k^i \hspace*{-0.7mm} + \hspace*{-0.7mm} 1}\Big)  \Big) \notag\\[-1mm]
&& \hspace*{-5mm} \ge R_{{\mathrm{PU}}_k}^{\mathrm{req}},\,\, \forall k, \quad  \quad \mbox{C4: } \overset{N_{\mathrm{F}}}{\underset{i=1}{\sum}} \Big( \tilde{q}_{\mathrm{ST}}^i + \overset{K}{\underset{k=1}{\sum}} \tilde{q}_{k}^i \Big) \le P_{\mathrm{max}}^{\mathrm{PT}}, \notag\\[-1mm]
&&\hspace*{-5mm}\mbox{C7, C8, C9: } \tilde{q}_k^i, \tilde{q}_{\mathrm{ST}}^i, \tilde{p}_{\mathrm{PU}_{k,j}}^i, \tilde{p}_{\mathrm{SU}_{k,j}}^i \ge 0, \forall i,k,j. \notag
\end{eqnarray}
Here, we note that the binary constraints C5 and C6 are absorbed in the objective function via the penalty terms $\tilde{I}_{\mathrm{PU}_{k,j}}^i$, $\tilde{I}_{\mathrm{SU}_{k,j}}^i$, and $\tilde{S}_{\mathrm{PU}_{k}}^i$ in \eqref{throughput-i-newdefine}. Due to the monotonicity of the objective function in  \eqref{pro-trans}, for the optimal solution, each subcarrier will be allocated to a single receiver in the primary network or a single pair of PU and SU in the secondary network, i.e., undesired co-channel interference is completely avoided.

For facilitating monotonic optimization, we define three auxiliary variables $u_{k,j}^i$, $v_{k,j}^i$, and $\xi_{k}^i$, which satisfy the following constraints:\vspace*{-2mm}
\begin{eqnarray}\label{constraint-MO}
\hspace*{-5mm}&&\hspace*{-7mm}\mbox{M1: }\hspace*{-0.5mm}  1 \hspace*{-1mm} \le \hspace*{-1mm} u_{k,j}^i \hspace*{-1mm} \le \hspace*{-1mm} 1 \hspace*{-0.7mm} + \hspace*{-0.7mm} \frac{\tilde{p}_{\mathrm{PU}_{k,j}}^i H_k^i}{ \tilde{p}_{\mathrm{SU}_{k,j}}^i H_k^i
\hspace*{-0.7mm} + \hspace*{-0.7mm} \tilde{I}_{\mathrm{PU}_{k,j}}^i  \hspace*{-0.7mm} + \hspace*{-0.7mm} 1}, \forall i,\hspace*{-0.2mm}j,\hspace*{-0.2mm} \forall k \hspace*{-1mm} \in\hspace*{-1mm} \mathcal{K}\hspace*{-0.2mm}(\hspace*{-0.2mm}i,j\hspace*{-0.2mm})\hspace*{-0.2mm}, \\[-1mm]
%%%%%
\hspace*{-5mm}&&\hspace*{-7mm}\mbox{M2: } \hspace*{-0.5mm} 1 \le v_{k,j}^i \le 1 \hspace*{-0.7mm} + \hspace*{-0.7mm} \frac{\tilde{p}_{\mathrm{SU}_{k,j}}^i \hspace*{-0.7mm} G_j^i}{\tilde{I}_{\mathrm{SU}_{k,j}}^i \hspace*{-0.7mm} + \hspace*{-0.7mm} 1},\forall i,j, \forall k\in\mathcal{K}\hspace*{-0.2mm}(\hspace*{-0.2mm}i,j\hspace*{-0.2mm}), \\[-1mm]
%%%%%
\hspace*{-5mm}&&\hspace*{-7mm}\mbox{M3: } \hspace*{-0.5mm} 1 \le \xi_{k}^i \le 1 \hspace*{-0.7mm} + \hspace*{-0.7mm} \frac{\tilde{q}_{k}^i F_k^i}{\tilde{S}_{\mathrm{PU}_{k}}^i \hspace*{-0.7mm} + \hspace*{-0.7mm} 1}, \forall i,k.
\end{eqnarray}
In addition, we define
\begin{eqnarray} \label{objective-MO}
\underline{\tilde{U}}^i \hspace*{-1.5mm} = \hspace*{-1.5mm} \sum_{j=1}^{J}  \hspace*{-0.5mm} \sum_{k \in  \mathcal{K}(i,j)} \hspace*{-1.5mm}  \frac{1}{2} [\log_2 \hspace*{-0.5mm} ( \hspace*{-0.5mm}u_{k,j}^i\hspace*{-0.5mm})^w \hspace*{-0.5mm} + \hspace*{-0.5mm}  \log_2\hspace*{-0.5mm} (\hspace*{-0.5mm}v_{k,j}^i\hspace*{-0.5mm})^\mu] \hspace*{-0.5mm} + \hspace*{-0.5mm} \sum_{k=1}^{K} \hspace*{-0.5mm}  \log_2\hspace*{-0.3mm} (\hspace*{-0.2mm}\xi_{k}^i\hspace*{-0.2mm})^w.
\end{eqnarray}
In fact, $\underline{\tilde{U}}^i$ serves as a lower bound of the original objective function $U^i$ in \eqref{throughput-i}. However, due to the monotonicity of the original problem in \eqref{pro}, we can obtain the optimal solution of \eqref{pro} by maximizing $\underline{\tilde{U}}^i$. In particular, the original problem in \eqref{pro} can be rewritten as a monotonic optimization problem in canonical form as follows:\vspace*{-2mm}
\begin{eqnarray} \label{pro-MO}
&&\hspace*{-6mm}\underset{\mathbf{u},\mathbf{v},\bm{\xi}, \bm{\varsigma}}{\maxo} \sum_{i=1}^{N_{\mathrm{F}}} \underline{\tilde{U}}^i  \quad \quad \mbox{s.t.}  (\mathbf{u},\mathbf{v},\bm{\xi}, \bm{\varsigma}) \in \mathcal{G}\cap\mathcal{H},
\end{eqnarray}
where vectors $\mathbf{u}$, $\mathbf{v}$, $\bm{\xi}$, and $\bm{\varsigma}$ contain all $u_{k,j}^i$, $v_{k,j}^i$, $\xi_{k}^i$, and $\varsigma_{k,j}^i$, respectively.
In \eqref{objective-MO}, $\mathcal{G}$ denotes a normal set \cite{zhang2013monotonic} which accounts for the limitation of the available radio resources. In particular, $\mathcal{G}$ is spanned by constraints $\mbox{C1}\mbox{a}$, C3, C4, C9, M1, M2, and M3.
Besides, $\mathcal{H}$ is a conormal set which is spanned by constraints C1b and C2. We note that the objective function and all constraint functions in \eqref{pro-MO} are monotonically increasing functions. Hence, from the theory of monotonic optimization, we know that the globally optimal solution of  optimization problem \eqref{pro-MO} can be obtained by employing the outer polyblock approximation approach. In particular, a globally optimal solution of \eqref{pro-MO} can be obtained via an outer polyblock approximation based algorithm as developed in our previous works \cite{sun2016optimalJournal,sun2017MISONOMA}, and the details are omitted here due to page limitation.

We note that the monotonic optimization-based resource allocation algorithm entails a high computational complexity as the search space of the optimal solution grows exponentially with the number of vertexes involved in the problem, e.g. $2N_{\mathrm{F}}(K+KJ)$ for the considered system.
Nevertheless, the performance of the optimal resource allocation policy serves as a benchmark for any suboptimal algorithm. In the next section, we develop a suboptimal scheme, which finds a locally optimal solution for problem \eqref{pro}, but requires only polynomial time computational complexity.

\vspace*{-1mm}
\subsection{Suboptimal Solution}%
To facilitate the design of the suboptimal algorithm, we rewrite the original objective function in \eqref{throughput-i} as\vspace*{-2mm}
\begin{eqnarray} \label{throughput-i-suboptimal}
\hspace*{-6mm} \overline{U}^i\hspace*{-3mm}
&=& \hspace*{-3mm}   \sum_{j=1}^{J}   \sum_{k\in \mathcal{K}(i,j)} \hspace*{-1mm}  \frac{1}{2} \Big[w  \log_2 \Big( 1 \hspace*{-0.7mm} + \hspace*{-0.7mm} \frac{\tilde{p}_{\mathrm{PU}_{k,j}}^i H_k^i}{ \tilde{p}_{\mathrm{SU}_{k,j}}^i H_k^i
\hspace*{-0.7mm} + \hspace*{-0.7mm} 1}\Big) \notag \\[-3mm]
%%%%%
\hspace*{-6mm} &+& \hspace*{-3mm} \hspace*{-0.7mm} \mu  \hspace*{-0.4mm}  \log_2 (1 \hspace*{-0.7mm} + \hspace*{-0.7mm} \tilde{p}_{\mathrm{SU}_{k,j}}^i \hspace*{-0.7mm} G_j^i) \Big] \hspace*{-0.7mm} + \hspace*{-0.7mm} \sum_{k=1}^{K} \hspace*{-0.7mm} w \log_2 \Big( 1 \hspace*{-0.7mm} + \hspace*{-0.7mm} \tilde{q}_{k}^i F_k^i\Big).
\end{eqnarray}
We note that the multiplicative terms in \eqref{throughput-i-suboptimal} involving binary variables and  power allocation variables, i.e., $\tilde{p}_{\mathrm{PU}_{k,j}}^i = s_{k,j}^i p_{\mathrm{PU}_k}^i$, $\tilde{p}_{\mathrm{SU}_{k,j}}^i = s_{k,j}^i p_{\mathrm{SU}_j}^i$, $\tilde{q}_{k}^i = c_k^i q_{k}^i$, and $\tilde{q}_{\mathrm{ST}}^i = c_{\mathrm{ST}}^i q_{\mathrm{ST}}^i$, are obstacles for the design of computationally efficient resource allocation algorithms. To handle this issue, we adopt the big-M method \cite{lee2011mixed} to decompose the multiplicative terms.
In particular, we impose the following additional constraints:\vspace*{-2mm}
\begin{eqnarray}
&&\hspace*{-12mm}
\mbox{C10: } \hspace*{-1mm} \tilde{q}_k^i  \le  P_{\mathrm{max}}^{\mathrm{PT}} c_{k}^i,\hspace*{20mm}
\mbox{C11: } \hspace*{-1mm} \tilde{q}_{\mathrm{ST}}^i  \le  P_{\mathrm{max}}^{\mathrm{PT}} c_{\mathrm{ST}}^i,
\\
%%%%%%%%
&&\hspace*{-12mm}
\mbox{C12: } \hspace*{-1mm} \tilde{p}_{\mathrm{PU}_{k,j}}^i  \le  P_{\mathrm{max}}^{\mathrm{ST}} s_{k,j}^i, \hspace*{11.5mm}
\mbox{C13: } \hspace*{-1mm} \tilde{p}_{\mathrm{PU}_{k,j}}^i \hspace*{-0.5mm} \le \hspace*{-1mm} P_{\mathrm{max}}^{\mathrm{ST}} s_{k,j}^i,\\
%%%%%%
&&\hspace*{-12mm}
\mbox{C14: }\hspace*{-0.7mm}  \tilde{q}_k^i  \ge  q_k^i  - (1  - c_k^i)  P_{\mathrm{max}}^{\mathrm{PT}}, \hspace*{8.2mm}
\mbox{C15: }\hspace*{-0.7mm} \tilde{q}_k^i  \le  q_k^i,
\\
%%%%%%
&&\hspace*{-12mm}
\mbox{C16: }\hspace*{-0.7mm}  \tilde{q}_{\mathrm{ST}}^i \hspace*{-0.5mm} \ge \hspace*{-0.8mm} q_{\mathrm{ST}}^i \hspace*{-1mm} - \hspace*{-1mm}(1 \hspace*{-1mm} - \hspace*{-0.5mm}c_{\mathrm{ST}}^i) \hspace*{-0.5mm} P_{\mathrm{max}}^{\mathrm{PT}}, \hspace*{7.5mm}
\mbox{C17: }\hspace*{-0.7mm} \tilde{q}_{\mathrm{ST}}^i  \le  q_{\mathrm{ST}}^i,
\\
%%%%%%%
&&\hspace*{-12mm}
\mbox{C18: }\hspace*{-0.7mm}  \tilde{p}_{\mathrm{PU}_{k,j}}^i \hspace*{-0.5mm} \ge \hspace*{-0.8mm} p_{\mathrm{PU}_{k}}^i \hspace*{-1mm} - \hspace*{-1mm}(1 \hspace*{-1mm} - \hspace*{-0.5mm}s_{k,j}^i) \hspace*{-0.5mm} P_{\mathrm{max}}^{\mathrm{ST}},  \hspace*{1mm}
\mbox{C19: }\hspace*{-0.7mm} \tilde{p}_{\mathrm{PU}_{k,j}}^i \hspace*{-0.5mm} \le \hspace*{-0.5mm} p_{\mathrm{PU}_{k}}^i,
\\
%%%%%%
&&\hspace*{-12mm}
\mbox{C20: }\hspace*{-0.7mm}  \tilde{p}_{\mathrm{SU}_{k,j}}^i  \ge \hspace*{-0.8mm} p_{\mathrm{SU}_{j}}^i \hspace*{-1mm} - \hspace*{-1mm}(1 \hspace*{-1mm} - \hspace*{-0.5mm}s_{k,j}^i) \hspace*{-0.5mm} P_{\mathrm{max}}^{\mathrm{ST}}, \hspace*{1.5mm}
\mbox{C21: }\hspace*{-0.7mm} \tilde{p}_{\mathrm{SU}_{k,j}}^i \hspace*{-0.5mm} \le \hspace*{-0.5mm} p_{\mathrm{SU}_{j}}^i.
\end{eqnarray}
Besides, in order to handle the binary constraints C5 and C6 in problem \eqref{pro}, we replace constraints C5 and C6 with their equivalent constraints $\{\mbox{C5a, C5b}\}$ and $\{\mbox{C6a, C6b, C6c}\}$, respectively, which are given by\vspace*{-2mm}
\begin{eqnarray}\label{integer-relax1}
&&\hspace*{-9mm}\text{C5}\mbox{a: } \hspace*{-1mm} \overset{{N_{\mathrm{F}}}}{\underset{i=1}{\sum}} \overset{K}{\underset{k=1}{\sum}}  \overset{J}{\underset{j=1}{\sum}} \hspace*{-0.5mm} s_{k,j}^i \hspace*{-0.5mm} - \hspace*{-0.5mm} (s_{k,j}^i)^2 \hspace*{-0.5mm} \le \hspace*{-0.5mm}  0, \,\, \,\,\,  \text{ C5}\mbox{b: } 0 \hspace*{-0.5mm} \le \hspace*{-0.5mm} s_{k,j}^i \hspace*{-0.5mm} \le \hspace*{-0.5mm} 1,  \\[-2mm]
%%%%%%%%%%%%%%%%%
\label{integer-relax2}
&&\hspace*{-9mm}\text{C6}\mbox{a: } \hspace*{-1mm} \overset{{N_{\mathrm{F}}}}{\underset{i=1}{\sum}} \hspace*{-0.5mm}\Big(c_{\mathrm{ST}}^i -(c_{\mathrm{ST}}^i)^2+ \overset{K}{\underset{k=1}{\sum}} \hspace*{-0.5mm} c_{k}^i \hspace*{-0.5mm} - \hspace*{-0.5mm} (c_{k}^i)^2 \Big)\hspace*{-0.5mm} \le \hspace*{-0.5mm}  0, \\
%%%%%%%%%%%%%%%%%%
\label{integer-relax3}
&&\hspace*{-9mm}\text{C6}\mbox{b: } 0 \hspace*{-0.5mm} \le \hspace*{-0.5mm} c_{\mathrm{ST}}^i \hspace*{-0.5mm} \le \hspace*{-0.5mm} 1, \quad \text{and} \quad
\text{C6}\mbox{c: } 0 \hspace*{-0.5mm} \le \hspace*{-0.5mm} c_{k}^i \hspace*{-0.5mm} \le \hspace*{-0.5mm} 1.
\end{eqnarray}
We note that in \eqref{integer-relax1} and \eqref{integer-relax3}, $s_{k,j}^i$, $c_{\mathrm{ST}}^i$, and $c_{k}^i$ are continuous in the interval between zero and one.
However, constraints C5a and C6a are reverse convex functions \cite{ng2015power} which makes problem \eqref{pro} still non-convex.
To resolve this issue, we reformulate the problem in \eqref{pro} as\vspace*{-2mm}
\begin{eqnarray} \label{penalty-pro}
\hspace*{-4mm}&&\hspace*{-5mm}\underset{\tilde{\mathbf{p}},\tilde{\mathbf{q}},\mathbf{c},\mathbf{s}}{\mino}\,\,  \sum_{i=1}^{{N_{\mathrm{F}}}}\Big[ - \overline{U}^i + \rho \Big( \overset{K}{\underset{k=1}{\sum}}  \overset{J}{\underset{j=1}{\sum}}  s_{k,j}^i  -  (s_{k,j}^i)^2\notag \\ [-2mm]
\hspace*{-4mm}&&\hspace*{11mm} + \overset{K}{\underset{k=1}{\sum}}  c_{k}^i  -  (c_{k}^i)^2 + c_{\mathrm{ST}}^i -(c_{\mathrm{ST}}^i)^2 \Big) \Big]\notag \\
\hspace*{-4mm}&&\hspace*{-4mm}\mbox{s.t.}  \hspace*{3mm}\,\,\mbox{C1--C4}, \mbox{C5b, C6b, C6c, C7--C21},
\end{eqnarray}
where the $\tilde{\mathbf{p}},\tilde{\mathbf{q}},\mathbf{c},\mathbf{s}$ contain all $\{\tilde{p}_{\mathrm{PU}_{k,j}}^i, \tilde{p}_{\mathrm{SU}_{k,j}}^i\}$, $\{\tilde{q}_k^i,\tilde{q}_{\mathrm{ST}}^i\}$, $\{c_{\mathrm{ST}}^i, c_{k}^i\}$, and $s_{k,j}^i$, respectively. Constant $\rho \gg 1$ is the penalty factor. In fact, $\rho \Big( \overset{K}{\underset{k=1}{\sum}}  \overset{J}{\underset{j=1}{\sum}}  s_{k,j}^i \hspace*{-0.5mm} - \hspace*{-0.5mm} (s_{k,j}^i)^2  \hspace*{-0.5mm}+ \hspace*{-0.5mm} \overset{K}{\underset{k=1}{\sum}}  c_{k}^i \hspace*{-0.5mm} - \hspace*{-0.5mm} (c_{k}^i)^2 + c_{\mathrm{ST}}^i -(c_{\mathrm{ST}}^i)^2  \Big)$ in \eqref{penalty-pro} is the penalty term that penalizes the violation of constraints C5a and C6a, which forces the optimization variables $s_{k,j}^i$, $c_{k}^i$, and $c_{\mathrm{ST}}^i$ to be zero or one. It is shown in \cite{sun2016optimalJournal,ng2015power} that \eqref{penalty-pro} and \eqref{pro} are equivalent for $\rho \gg 1$.

\begin{table} \vspace*{-1mm}
%\caption{Iterative Resource Allocation Algorithm for Suboptimal Online and Optimal Offline Designs.}\label{table:algorithm}\vspace*{-0cm}
%\small
\begin{algorithm} [H]                    % enter the algorithm environment
\caption{Successive Convex Approximation}          % give the algorithm a caption
\label{alg1}                           % and a label for \ref{} commands later in the document
\begin{algorithmic} [1]
\small          % enter the algorithmic environment
\STATE Initialize the penalty factor $\rho \hspace*{-0.8mm} \gg \hspace*{-0.8mm} 1$, iteration index $r=1$, and initial point $\tilde{\mathbf{p}}^{(1)}$, $\tilde{\mathbf{q}}^{(1)}$, $\mathbf{c}^{(1)}$, and $\mathbf{s}^{(1)}$

\REPEAT
%%%%%%%%%%%%%%%%%%%%
\STATE Solve \eqref{dc} for given $\tilde{\mathbf{p}}^{(r)}$, $\tilde{\mathbf{q}}^{(r)}$, $\mathbf{c}^{(r)}$, and $\mathbf{s}^{(r)}$ and store the intermediate resource allocation policy $\{\tilde{\mathbf{p}}, \tilde{\mathbf{q}}, \mathbf{c}, \mathbf{s}\}$

\STATE Set $r=r+1$ and $\tilde{\mathbf{p}}^{(r)}=\tilde{\mathbf{p}}$, $\tilde{\mathbf{q}}^{(r)}=\tilde{\mathbf{q}}$, $\mathbf{c}^{(r)}=\mathbf{c}$, and $\mathbf{s}^{(r)}=\mathbf{s}$

\UNTIL convergence

\STATE Obtain final resource allocation policy $\tilde{\mathbf{p}}^{*}=\tilde{\mathbf{p}}^{(r)}$, $\tilde{\mathbf{q}}^{*}=\tilde{\mathbf{q}}^{(r)}$, $\mathbf{c}^{*}=\mathbf{c}^{(r)}$, and $\mathbf{s}^{*}=\mathbf{s}^{(r)}$

\end{algorithmic}
\end{algorithm}\vspace*{-8mm}
\end{table}

Now, the remaining non-convexity of problem \eqref{penalty-pro} is due to constraints $\mbox{C1}$, $\mbox{C2}$, and the objective function. However,  \eqref{penalty-pro} can be rewritten in form of a standard  difference of convex (d.c.) programming problem \cite{ng2015power} as: \vspace*{-0mm}
\begin{eqnarray}\label{dc-penalty-pro}
\hspace*{-5mm}&&\hspace*{-0mm}\underset{\tilde{\mathbf{p}},\tilde{\mathbf{q}},\mathbf{c},\mathbf{s}}{\mino}\,\, \,\, \bm{F}(\tilde{\mathbf{p}},\tilde{\mathbf{q}})-\bm{G}(\tilde{\mathbf{p}},\tilde{\mathbf{q}})+\rho(\bm{H}(\mathbf{c},\mathbf{s})-\bm{M}(\mathbf{c},\mathbf{s})) \notag \\[-0.5mm]
\hspace*{-5mm}\mbox{s.t.}\hspace*{-8mm} &&\hspace*{4mm}
\widetilde{\mbox{C1}}\mbox{: }   \bm{B}_{k,j}^i(\tilde{\mathbf{p}},\tilde{\mathbf{q}}) - \bm{D}_{k,j}^i(\tilde{\mathbf{p}},\tilde{\mathbf{q}}) \le 0,  \notag \\[-0.5mm]
%%%%%%%%
\hspace*{-5mm}&&\hspace*{4mm}
\widetilde{\mbox{C2}}\mbox{: }    \bm{R}_{k,j}^i(\tilde{\mathbf{p}},\tilde{\mathbf{q}}) - \bm{T}_{k,j}^i(\tilde{\mathbf{p}},\tilde{\mathbf{q}}) \le 0,  \notag \\[-0.5mm]
\hspace*{-5mm}&&\hspace*{4mm} \mbox{C3, C4, C5b, C6b, C6c, C7--C21, }
\end{eqnarray}
where \vspace*{-2mm}
\begin{eqnarray}
\hspace*{-2mm}\bm{F} (\tilde{\mathbf{p}},\tilde{\mathbf{q}}) \hspace*{-3mm}&=&\hspace*{-3mm} \sum_{i=1}^{{N_{\mathrm{F}}}} \hspace*{-1mm} \Big(\sum_{j=1}^{J}   \sum_{k\in \mathcal{K}(i,j)} \hspace*{-1mm}  \frac{1}{2} \big[w  \log_2 ( 1 \hspace*{-0.7mm} + \hspace*{-0.7mm} (\tilde{p}_{\mathrm{PU}_{k,j}}^i \hspace*{-1.3mm} + \hspace*{-0.7mm} \tilde{p}_{\mathrm{SU}_{k,j}}^i) H_k^i ) \notag \\[-3mm]
%%%
\hspace*{-2mm} &+& \hspace*{-3mm} \hspace*{-0.7mm} \mu  \hspace*{-0.4mm}  \log_2 (1 \hspace*{-0.7mm} + \hspace*{-0.7mm} \tilde{p}_{\mathrm{SU}_{k,j}}^i \hspace*{-0.7mm} G_j^i) \big] \hspace*{-0.7mm} + \hspace*{-0.7mm} \sum_{k=1}^{K} \hspace*{-0.7mm} w \log_2 ( 1 \hspace*{-0.7mm} + \hspace*{-0.7mm} \tilde{q}_{k}^i F_k^i) \Big),  \\[-3mm]
%%%%%%%%%%%%%%%%
\hspace*{-2mm}\bm{G}(\tilde{\mathbf{p}},\tilde{\mathbf{q}}) \hspace*{-3mm}&=&\hspace*{-3mm} \sum_{i=1}^{{N_{\mathrm{F}}}}  \sum_{j=1}^{J}   \sum_{k\in \mathcal{K}(i,j)} \hspace*{-1mm}  \frac{1}{2} w  \log_2 ( 1 \hspace*{-0.7mm} + \hspace*{-0.7mm}  \tilde{p}_{\mathrm{SU}_{k,j}}^i H_k^i ),  \\[-1mm]
%%%%%%%%%%%%%%%%
\hspace*{-2mm}\bm{H} (\mathbf{c},\mathbf{s})\hspace*{-3mm}&=&\hspace*{-3mm} \sum_{i=1}^{{N_{\mathrm{F}}}}\Big(  c_{\mathrm{ST}}^i  +  \overset{K}{\underset{k=1}{\sum}}  c_{k}^i +  \overset{K}{\underset{k=1}{\sum}}  \overset{J}{\underset{j=1}{\sum}}  s_{k,j}^i \Big), \,\,\,\, \text{and}\\[-1mm]
%%%%%%%%%%%%%%%%
\hspace*{-2mm}\bm{M} (\mathbf{c},\mathbf{s})\hspace*{-3mm}&=&\hspace*{-3mm} \sum_{i=1}^{{N_{\mathrm{F}}}}\Big(  (c_{\mathrm{ST}}^i)^2  +  \overset{K}{\underset{k=1}{\sum}}  (c_{k}^i)^2 +  \overset{K}{\underset{k=1}{\sum}}  \overset{J}{\underset{j=1}{\sum}}  (s_{k,j}^i)^2 \Big).
%%%%%%%%%%%%%%%%%
\end{eqnarray}
The definitions of $\bm{B}_{k,j}^i(\tilde{\mathbf{p}},\tilde{\mathbf{q}})$ and $\bm{D}_{k,j}^i(\tilde{\mathbf{p}},\tilde{\mathbf{q}})$ in $\widetilde{\mbox{C1}}$, and the definitions of $\bm{R}_{k,j}^i(\tilde{\mathbf{p}},\tilde{\mathbf{q}})$ and $\bm{T}_{k,j}^i(\tilde{\mathbf{p}},\tilde{\mathbf{q}})$ in $\widetilde{\mbox{C2}}$ are similar to those of $\bm{F} (\tilde{\mathbf{p}},\tilde{\mathbf{q}})$ and $\bm{G} (\tilde{\mathbf{p}},\tilde{\mathbf{q}})$. In particular, constraints $\widetilde{\mbox{C1}}$ and $\widetilde{\mbox{C2}}$ are written as a difference of logarithmic functions.
We note that the problems in \eqref{pro-trans} and \eqref{dc-penalty-pro} are equivalent in the sense that they have the same optimal solution.
Thus, we can obtain a locally optimal solution of \eqref{dc-penalty-pro} by applying successive convex approximation \cite{dinh2010local}.
In particular, for any feasible point $\tilde{\mathbf{p}}^{(r)}$, $\tilde{\mathbf{q}}^{(r)}$, $\mathbf{c}^{(r)}$, and $\mathbf{s}^{(r)}$, we have the following inequalities:\vspace*{-0mm}
\begin{eqnarray}\label{ineq1}
\hspace*{-6mm} \bm{G}(\tilde{\mathbf{p}},\tilde{\mathbf{q}})
\hspace*{-2mm} &\ge& \hspace*{-2mm} \bm{G}(\tilde{\mathbf{p}}^{(r)},\tilde{\mathbf{q}}^{(r)})  +  \Tr(\nabla_{\tilde{\mathbf{p}}} \bm{G}(\tilde{\mathbf{p}},\tilde{\mathbf{q}})(\tilde{\mathbf{p}} \hspace*{-0mm} - \hspace*{-0mm}\tilde{\mathbf{p}}^{(r)}) ) \notag \\[-1mm]
%%%%%%%%
\hspace*{-2mm} &+& \hspace*{-2mm} \Tr(\nabla_{\tilde{\mathbf{q}}} \bm{G}(\tilde{\mathbf{p}},\tilde{\mathbf{q}})(\tilde{\mathbf{q}} \hspace*{-0mm} - \hspace*{-0mm}\tilde{\mathbf{q}}^{(r)}) ) \notag\\
\hspace*{-2mm} &\triangleq& \hspace*{-2mm}\overline{G} (\tilde{\mathbf{p}},\tilde{\mathbf{q}}, \tilde{\mathbf{p}}^{(r)},\tilde{\mathbf{q}}^{(r)}),
\end{eqnarray}
where the right hand side of \eqref{ineq1} is an affine function and represents a global underestimation of $\bm{G}(\tilde{\mathbf{p}},\tilde{\mathbf{q}})$. Similarly, we denote $\overline{\bm{M}}(\mathbf{c},\mathbf{s}, \mathbf{c}^{(r)},\mathbf{s}^{(r)})$, $\overline{\bm{D}}_{k,j}^i(\tilde{\mathbf{p}},\tilde{\mathbf{q}}, \tilde{\mathbf{p}}^{(r)},\tilde{\mathbf{q}}^{(r)})$, and $\overline{\bm{T}}_{k,j}^i(\tilde{\mathbf{p}},\tilde{\mathbf{q}}, \tilde{\mathbf{p}}^{(r)},\tilde{\mathbf{q}}^{(r)})$ as the global underestimations of $\bm{M} (\mathbf{c},\mathbf{s})$, $\bm{D}_{k,j}^i(\tilde{\mathbf{p}},\tilde{\mathbf{q}})$, and $\bm{T}_{k,j}^i(\tilde{\mathbf{p}},\tilde{\mathbf{q}})$, respectively.

Therefore, for any given $\tilde{\mathbf{p}}^{(r)}$, $\tilde{\mathbf{q}}^{(r)}$, $\mathbf{c}^{(r)}$, and $\mathbf{s}^{(r)}$, we can obtain a lower bound of \eqref{dc-penalty-pro} by solving the following optimization problem: \vspace*{-3mm}
\begin{eqnarray}\label{dc}
\hspace*{-5mm}&&\hspace*{-0mm}\underset{\tilde{\mathbf{p}},\tilde{\mathbf{q}},\mathbf{c},\mathbf{s}}{\mino}\,\, \,\, \bm{F}(\tilde{\mathbf{p}},\tilde{\mathbf{q}})- \overline{\bm{G}}(\tilde{\mathbf{p}},\tilde{\mathbf{q}},\tilde{\mathbf{p}}^{(r)},\tilde{\mathbf{q}}^{(r)}) \notag \\ [-2mm]
&&\hspace*{15mm} +\rho(\bm{H}(\mathbf{c},\mathbf{s})- \overline{\bm{M}}(\mathbf{c},\mathbf{s}, \mathbf{c}^{(r)},\mathbf{s}^{(r)})) \notag \\
\hspace*{-5mm}\mbox{s.t.}\hspace*{-8mm} &&\hspace*{4mm}
\widetilde{\mbox{C1}}\mbox{: }   \bm{B}_{k,j}^i(\tilde{\mathbf{p}},\tilde{\mathbf{q}}) - \overline{\bm{D}}_{k,j}^i(\tilde{\mathbf{p}},\tilde{\mathbf{q}}, \tilde{\mathbf{p}}^{(r)},\tilde{\mathbf{q}}^{(r)}) \le 0,  \notag \\
%%%%%%%%
\hspace*{-5mm}&&\hspace*{4mm}
\widetilde{\mbox{C2}}\mbox{: }    \bm{R}_{k,j}^i(\tilde{\mathbf{p}},\tilde{\mathbf{q}}) - \overline{\bm{T}}_{k,j}^i(\tilde{\mathbf{p}},\tilde{\mathbf{q}}, \tilde{\mathbf{p}}^{(r)},\tilde{\mathbf{q}}^{(r)}) \le 0,  \notag \\
\hspace*{-5mm}&&\hspace*{4mm} \mbox{C3, C4, C5b, C6b, C6c, C7--C21. }
\end{eqnarray}
We successively tighten the obtained lower bound by applying the iterative algorithm summarized in \textbf{Algorithm 1}.
The proposed suboptimal iterative algorithm converges to a locally optimal solution of \eqref{dc-penalty-pro} with polynomial time computational complexity \cite{dinh2010local}.
We note that the convex problem in \eqref{dc} can be solved efficiently by standard convex program solvers such as CVX \cite{website:CVX}.

\vspace*{-2mm}
\section{Simulation Results}%
\begin{table}[t]\vspace*{-0mm}\caption{System parameters}\vspace*{-2mm}\label{tab:parameters} %\vspace*{-1mm}
\newcommand{\tabincell}[2]{\begin{tabular}{@{}#1@{}}#2\end{tabular}}
\centering
\begin{tabular}{|l|l|}\hline
\hspace*{-1mm}Carrier center frequency and bandwidth & $2$ GHz and $2.5$ MHz \\
\hline
\hspace*{-1mm}Number of subcarriers, ${N_{\mathrm{F}}}$ & $32$  \\
\hline
\hspace*{-1mm}Bandwidth of subcarrier  & $78$ kHz  \\
\hline
\hspace*{-1mm}Primary and secondary BS antenna gain & \mbox{$10$ dBi} and \mbox{$5$ dBi}\\
\hline
\hspace*{-1mm}Path loss exponent and reference distance, $D_{\mathrm{ref}}$   &  \mbox{$3.6$} and  $10$ meters  \\
\hline
\hspace*{-1mm}Receiver noise power, $\sigma_{{\mathrm{ST}}}^2, \sigma_{{\mathrm{PU}}_k}^2, \sigma_{{\mathrm{SU}}_j}^2$   &   \mbox{$-110$ dBm}  \\
\hline
\hspace*{-1mm}Maximum  power at primary BS, $P_{\mathrm{max}}^{\mathrm{PT}}$ &  \mbox{$40$ dBm}   \\
\hline
\hspace*{-1mm}Maximum  power at secondary BS, $P_{\mathrm{max}}^{\mathrm{ST}}$ &  \mbox{$40$ dBm}   \\
\hline
\hspace*{-1mm}Minimum required rate for PUs, $R_{{\mathrm{PU}}_k}^{\mathrm{req}}=R_{\mathrm{req}}$   &  $1$ bits/s/Hz  \\
\hline
\hspace*{-1mm}Penalty factor $\rho$ for \textbf{Algorithm 1} &  $10 \hspace*{-0.5mm} \log_2(1 \hspace*{-1mm} +\hspace*{-1mm} P_{\mathrm{max}}\hspace*{-0.5mm}/\hspace*{-0.5mm}\sigma_{\mathrm{ST}}^2)$  \\
\hline
\end{tabular}
\vspace*{-3mm}
\end{table}
In this section, we evaluate the system performance of the proposed scheme via simulations. The adopted simulation parameters are given in Table \ref{tab:parameters}, unless specified otherwise.
We assume that the primary BS is $L$ meters away from the secondary BS.
There are $K$ PUs and $J$ SUs which are randomly and uniformly distributed between the reference distance and the maximum service distances of $D_{\mathrm{PT}}=500$ meters and $D_{\mathrm{ST}}=150$ meters for the primary BS and the secondary BS, respectively.
The small-scale fading of the primary BS-to-PU channels, the secondary BS-to-PU channels, the secondary BS-to-SU channels, and the link between the primary BS and the secondary BS is modeled as independent and identically Rayleigh distributed. The weights of the PUs and SUs are set as $w=2$ and $\mu=1$, respectively, to provide higher priority for maximizing the throughput of the PUs. The results shown in this section are averaged over different realizations of pathloss and multipath fading.

For comparison, we also consider the performance of two baseline schemes.
For baseline scheme $1$, we consider a traditional multicarrier cognitive relaying system where the SUs cannot perform SIC for cancelling the interference from PUs in the secondary network.
For baseline scheme $2$, the user pair on each subcarrier in the secondary network is  selected randomly and we optimize the corresponding transmit powers for the PUs and SUs.

\vspace*{-2mm}
\subsection{Average User Throughput vs. Normalized Distance}
\begin{figure}
\centering\vspace*{-0mm}
\includegraphics[width=3.45in]{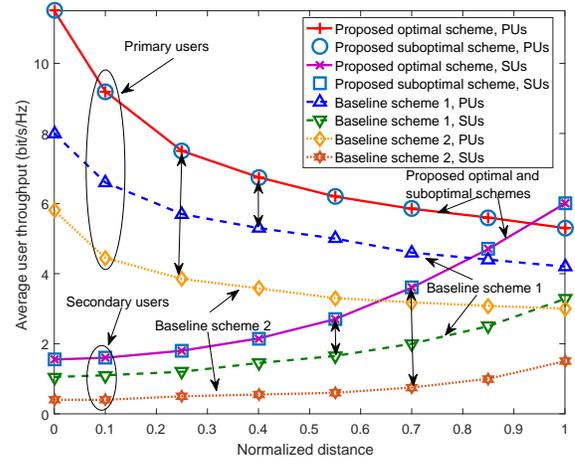}\vspace*{-3mm}
\caption{Average user data rate (bit/s/Hz) versus the normalized distance between the primary and secondary BSs for different resource allocation schemes. The double-sided arrows indicate the performance gains of the proposed optimal scheme compared to the baseline schemes.}
\label{fig:data-rate-distance}\vspace*{-5mm}
\end{figure}

In Figure \ref{fig:data-rate-distance}, we investigate the average user throughput of all PUs and all SUs versus (vs.) the normalized distance between the primary BS and the secondary BS, for $K=3$ PUs and $J=3$ SUs.
In particular, the normalized distance between the primary BS and the secondary BS is given by $\frac{L-D_{\mathrm{ref}}}{D_{\mathrm{PT}} - D_{\mathrm{ref}}}$.
As can be observed from Figure \ref{fig:data-rate-distance}, for the proposed optimal and suboptimal schemes, the average user throughput of the PUs  decreases monotonically with the normalized distance. In particular, as the normalized distance increases, the quality of the primary BS-to-secondary BS link deteriorates and becomes the bottleneck for the throughput of the assisted PUs.
Thus, the primary BS is more reluctant to let the secondary BS assist in the information transmission since forwarding the information over a weak channel requires high transmit power.
As a result, the potential throughput gain of the PUs introduced by the secondary BS serving as a relay diminishes with increasing distance between the primary and secondary BSs.
%%%%%
On the other hand, it can be observed that the average user throughput of the SUs increases with the normalized distance.
In particular, as the normalized distance increases, the secondary BS assists a decreasing number of PUs since the quality of the information forwarding channel between the primary BS and the secondary BS becomes worse.
Thus, less power is used at the secondary BS for satisfying the QoS requirements of the assisted PUs  and the newly available power can be reallocated to SUs for improving the system throughput.
We also note that the proposed suboptimal scheme closely approaches the performance of the proposed optimal resource allocation scheme.
On the other hand, Figure \ref{fig:data-rate-distance} shows that both baseline schemes achieve a substantially lower average throughput for the PUs and the SUs compared to the proposed schemes. In particular, since baseline scheme 1 does not utilize interference cancellation at the receivers, the co-channel interference between the PU and the SU  in the secondary network degrades the average user throughput of both the PUs and the SUs. Besides, for baseline scheme 2, the non-optimality of the subcarrier allocation leads to a substantial reduction in the average user throughputs. In particular, for a normalized distance of $0.55$, the proposed schemes can achieve roughly a $24\%$ and $87\%$  higher average user throughput for the PUs and a $64\%$ and $280\%$ higher average user throughput for the SUs than baseline schemes $1$ and $2$, respectively.

\vspace*{-1mm}
\subsection{Average User System Throughput vs. Number of Users}
\begin{figure}
\centering\vspace*{-0mm}
\includegraphics[width=3.45in]{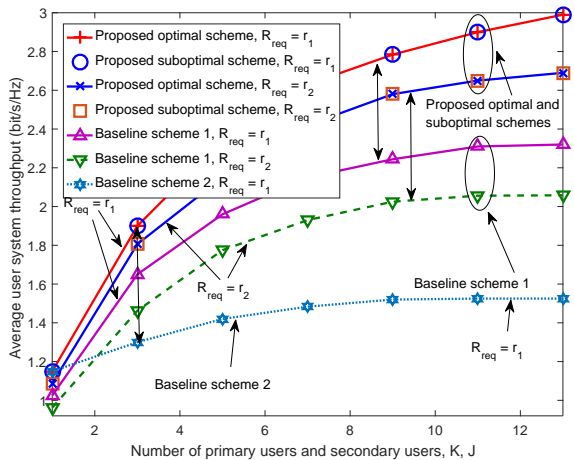}\vspace*{-3mm}
\caption{Average user system  throughput (bit/s/Hz) versus the number of primary and secondary users, $K$, $J$, for different resource allocation schemes. The double-sided arrows indicate the performance gains of the proposed optimal scheme compared to the baseline schemes.}
\label{fig:throughput-numuser}\vspace*{-5mm}
\end{figure}
In Figure \ref{fig:throughput-numuser}, we investigate the average user system throughput vs. the number of PUs and SUs for a normalized distance between the primary and secondary BSs of $0.6$ and different minimum QoS requirements for the PUs, i.e., $r_1 = 1$ bits/s/Hz and $r_2 = 2$ bits/s/Hz. We assume that the numbers of PUs and SUs are identical, i.e., $K=J$. The average user system throughput is calculated as $ \frac{ \sum_{i=1}^{N_{\mathrm{F}}} \sum_{k=1}^{K} (c_k^i  C_{\mathrm{PU}_{k}}^i + \sum_{j=1}^{J} s_{k,j}^i (R_{\mathrm{PU}_{k,j}}^i+R_{\mathrm{SU}_{k,j}}^i))}{K+J}$.
As can be seen from Figure \ref{fig:throughput-numuser}, the average user system throughput of both proposed schemes and the baseline schemes increase monotonically with the number of users due to the ability of these schemes to exploit multiuser diversity. Nevertheless, it can be observed that the average user system throughput of the proposed schemes grows faster with increasing number of users than that of the baseline schemes. In fact, compared to baseline scheme 1, since the proposed schemes utilize SIC for multiuser detection at the SUs, the co-channel interference  is significantly reduced in the secondary network. Besides, the proposed schemes exploit the power domain to facilitate  multiuser access which offers additional degrees of freedom for user scheduling and power allocation. Moreover, since baseline scheme 2 adopts random user allocation in the secondary network, it can only exploit the multiuser diversity of the PUs to improve the average user throughput, which leads to a marginal performance gain.
On the other hand, both the proposed schemes and the baseline schemes achieve a lower average user system throughput for larger minimum data rate requirements.
In fact, both the primary BS and the secondary BS have to allocate more power and frequency resources to the PUs to meet more stringent QoS requirements even if the conditions of the corresponding channels are poor.
We note that the proposed suboptimal scheme achieves a similar performance as the proposed optimal scheme in all considered scenarios  but entails a lower computational complexity.

\vspace*{-1mm}
\section{Conclusion}
\vspace*{-0mm}

In this paper, we studied the power and subcarrier allocation design for cooperative cognitive relaying MC-NOMA systems. The resource allocation design was formulated as a mixed-integer non-convex optimization problem for the maximization of the weighted system throughout. The optimal resource allocation policy was obtained by optimally solving the formulated problem via monotonic optimization. Besides, we developed a low-complexity suboptimal scheme for finding a locally optimal solution in order to achieve a balance between optimality and computational complexity. Our simulation results unveiled that the proposed cognitive relaying MC-NOMA system achieves a significantly higher system throughput compared to traditional multicarrier cognitive relaying systems.

\vspace*{-1mm}
\bibliographystyle{IEEEtran}
\bibliography{NOMA_cognitive_relaying}

\end{document}